# Do Earthquakes Wake More Slopes Than They Calm?

Ashok Dahal[1], Hakan Tanyaş[1], Alexander L. Handwerger[2,3], Luigi Lombardo[1], Eric Fielding[2]

## Abstract

Strong earthquakes can have a dual impact on the slow-moving landslide budget of a landscape by either triggering new actively deforming hillslopes or stabilizing existing ones. However, their regional-scale influence on hillslope processes remains poorly understood due to limited observational coverage and the masking effects of long-wavelength post-seismic signals. This knowledge gap constrains our ability to assess post-seismic landslide hazards, particularly since actively deforming hillslopes may evolve into rapid failures, especially under the influence of meteorological factors following a major earthquake. This study addresses this gap by analyzing slope instability across approximately 66,800 km$^2$ of southeastern Türkiye in the aftermath of the February 6, 2023, Kahramanmaraş earthquake sequence. We used Sentinel-1 InSAR time series spanning three years before and one year after the earthquakes and applied a convolution-based filtering method to isolate localized surface deformation from broader tectonic signals. Given the challenge of validating deformation anomalies at a regional scale without in-situ measurements, we cross-checked these locations using high-resolution optical imagery and assigned confidence levels based on observable surface features indicative of slow-moving landslides. The resulting inventories were further enriched by expert-based assessments. Our results show that the total area affected by slow-moving landslides increased from 12.36 km$^2$ to 159.38 km$^2$, reflecting a more than 12-fold rise. While a substantial number of slopes were newly activated (348), others became stabilized (28) or shifted to different deformation states (20). These findings highlight the complex response of hillslopes to seismic forcing and emphasize the importance of long-term monitoring to capture the full spectrum of earthquake-induced landslide behavior.

---

[1] University of Twente, Faculty of Geo-Information Science and Earth Observation (ITC), PO Box 217, Enschede, AE, 7500, Netherlands

[2] Jet Propulsion Laboratory, California Institute of Technology, Pasadena, CA, 91109, USA

[3] Joint Institute for Regional Earth System Science and Engineering, University of California, Los Angeles, Los Angeles, CA, 90095, USA

## 1. Introduction

Strong earthquakes (e.g., > $M_w$ 6.0) not only trigger widespread landslides during the coseismic phase but also disturb hillslopes, creating damage that weakens hillslope material, as well as fractures and fissures that enhance water infiltration leading to a reduction in effective normal stress and, consequently, frictional resistance (Bontemps et al., 2020; Gischig et al., 2016). As a result, the post-seismic period may witness increased or accelerated landslide activity.

Changes in hillslope conditions in the aftermath of strong earthquakes can contribute to the initiation, reactivation, or acceleration of slow-moving landslides (e.g., Bontemps et al., 2018; Carey et al., 2019; Cheaib et al., 2022; Lacroix et al., 2020, 2015), which are defined as coherent masses of soil and rock undergoing persistent slow (< 1 m/yr) downslope deformation (Akbarimehr et al., 2013; Handwerger et al., 2015).

However, earthquakes can also have opposing effects on slow-moving landslides. They may trigger the collapse of already deforming hillslopes or, under certain conditions, contribute to their temporary or long-term stabilization (He et al., 2023; Sadhasivam et al., 2024). While failure can occur on any hillslope where ground acceleration exceeds the critical threshold (Newmark, 1965), stabilization is rarer and may occur under specific conditions, such as, aftershock-induced densification of unconsolidated, loose and dry materials or internal stress redistribution (Lv et al., 2025; Song et al., 2022). Following seismic shaking, some landslides decelerate or return to their pre-earthquake movement rates, particularly when pre-event activity was low or when slope geometries favored stress relaxation (Kohler and Puzrin, 2022; Lv et al., 2025; Song et al., 2022).

As summarized above, the literature suggests that strong earthquakes can have a dual impact on the slow-moving landslide budget of a landscape. However, this impact at regional scales remains largely unexplored (Cheaib et al., 2022; He et al., 2023). Consequently, it is still unclear whether strong earthquakes ultimately increase or decrease the spatial extent of landscapes affected by slow-moving landslides (Lu et al., 2025). Moreover, the lack of a slow-moving landslide inventory for the post-seismic period leads to incomplete hazard assessments, as some newly formed or accelerated hillslope deformations may evolve into rapid failures during this time (Fan et al., 2018).

One key challenge is the limited availability of in-situ ground deformation measurements in large, mountainous areas affected by earthquakes (Bekaert et al., 2020). In such settings, spaceborne synthetic aperture radar interferometry (InSAR) is commonly used to detect surface deformation (Akbari and Motagh, 2012; Vassileva et al., 2023) and identify slow-moving landslides (e.g., Cai et al., 2022; Cao et al., 2023; He et al., 2023). However, post-seismic relaxation processes can introduce significant uncertainty in interpreting surface deformation (e.g., Liu et al., 2025; Yu et al., 2020), particularly when focusing on slow-moving landslides. Mechanisms such as fault afterslip, viscoelastic relaxation, and poroelastic rebound can all contribute to post-seismic deformation (Barbot and Fialko, 2010), potentially introducing broad-scale background signals that complicate the detection and attribution of localized, landslide-related displacements.

To isolate landslide deformation from tectonic and atmospheric signals, Bekaert et al (2020) implement a spatial double-difference technique. They apply two spatial filters to each interferogram, a small-radius kernel (100-200 m) to capture localized deformation and a larger-radius kernel (1-2 km) to estimate and remove long-wavelength signals associated with tectonic deformation or atmospheric delays. The difference between these filtered results yields a residual velocity map that highlights localized deformation (i.e., slow-moving landslides or other local scale deformation) while suppressing regional noise. However, this approach was applied to a single basin affected by the 2015 Gorkha earthquake, and thus, did not capture the broader regional evolution of slow-moving landslides following the earthquake.

In this study, we aim to map slow-moving landslides and associated hillslope deformation in the region affected by the 2023 Türkiye earthquake sequence using open-access standardized Sentinel-1 interferograms automatically processed by the JPL-Caltech Advanced Rapid Imaging and Analysis (ARIA) Center for Natural Hazards project (Bekaert et al., 2019). A key methodological objective is to remove long-wavelength post-seismic deformation signals that can obscure the localized displacements typically associated with slow-moving landslides. To achieve this, we implement a filtering approach that isolates short-wavelength signals, thereby enhancing the detectability of landslide-related deformation.

We complement our analysis with high-resolution optical imagery. This is used to: (i) minimize likely false positives, (ii) manually delineate slow-moving landslide polygons, (iii) assign confidence levels based indicative surface features of slow-moving landslides observable in the optical data, and (iv) label each location with the exposed elements within the identified polygons, including settlements, agricultural areas, roads, and reservoirs. As a result, we generate the most likely locations for slow-moving landslides.

In addition to spatial mapping, we analyze the temporal evolution of deformation through time series analysis, enabling the identification of trends, acceleration phases, and potential stabilization across different slopes.

Overall, our approach provides a quantitative assessment of changes in the slow-moving landslide budget by comparing the number and area of actively deforming slopes before and after the 2023 Türkiye earthquake sequence. It also offers complementary insights to support post-seismic landslide hazard assessment practices.

**2. Study Area**

Our study focuses on the region affected by the February 6, 2023, earthquake sequence, which primarily impacted southeastern Türkiye (**Figure** 1). The area enclosed by the 0.12g peak ground acceleration (PGA) isoline was previously surveyed for coseismic landslides (Görüm et al., 2023), however, slow-moving landslides within this zone have not yet been systematically investigated.

The study area is situated along the East Anatolian Fault (EAF), a major intracontinental left-lateral strike-slip fault zone that accommodates the relative motion between the Anatolian and Arabian plates (Arpat and Şaroğlu, 1972; Duman and Emre, 2013; Reilinger et al., 2006). The EAF extends from the Karlıova triple junction in the northeast, where it intersects the North Anatolian Fault (NAF), to the Iskenderun Gulf in the southwest, spanning approximately 550 to 700 km in length (Herece, 2008; Şengör and Yazıcı, 2020). This fault system comprises over 15 segments, with slip rates ranging from approximately 10 mm/yr in the north to around 4.5 mm/yr in the south (Aktug et al., 2016; Reilinger et al., 2006).

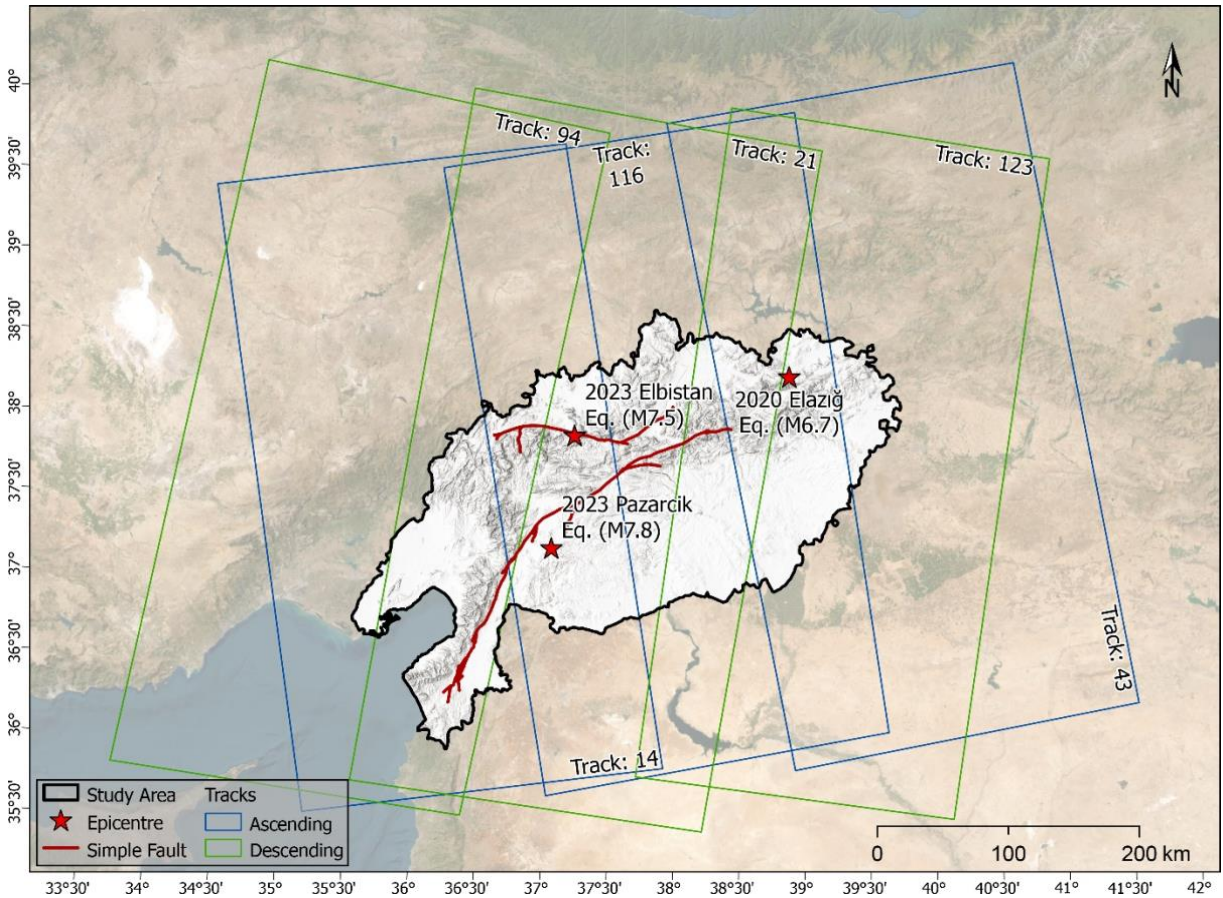

*Figure 1. Study area showing the mapping extent, selected Sentinel-1 tracks with frame coverage, and the epicenters of the February 6, 2023, earthquakes. Fault ruptures are from Reitman et al. (2023), and earthquake epicenters are provided by the U.S. Geological Survey.*

On February 6, 2023, this fault system experienced a devastating doublet earthquake sequence. The first event, a $M_w$ 7.8 earthquake, was nucleated near Pazarcık and was followed approximately nine hours later by a $M_w$ 7.5 earthquake near Ekinözü (Karabacak et al., 2023; Kurcer et al., 2023). Combined, the two mainshocks ruptured several segments of the East Anatolian Fault, producing surface ruptures of approximately 310 km for the $M_w$ 7.8 event and 140 km for the $M_w$ 7.6 event (Provost et al., 2024). Maximum coseismic slip reached about 10 meters on the East Anatolian and Sürgü faults (Abdelmeguid et al., 2023). InSAR and GPS analyses revealed pronounced lateral variability in slip magnitude and locking depth, highlighting the complex and heterogeneous rupture behavior along the fault (Li et al., 2023). Dynamic rupture modeling has shown that supershear rupture propagation occurred along several fault segments, especially in the southwestern portions of the EAF and along the Narlı Fault (Mai et al., 2023; Okuwaki et al., 2023). Owing to these characteristics, the February 6, 2023, earthquake doublet in southeastern Türkiye represents the most complex rupture evolution documented in the country over the past century, featuring backward branching and cascading multi-fault rupture (Liu et al., 2023).

Climatically, the study area lies within a transitional zone between the Mediterranean and continental regimes. The region experiences mild, wet winters and hot, dry summers, although significant spatial variation exists due to differences in elevation and latitude (Lionello et al., 2006). According to the updated Köppen-Geiger classification, the area spans several climate types, including cold semi-arid in the northern interior and hot-summer and warm-summer Mediterranean climates in the southern parts (Peel et al., 2007).

The region is also defined by highly variable topography, with elevation increasing markedly toward the north and east. Rugged mountain ranges, including parts of the Taurus and Eastern Anatolian mountains, dominate the landscape. These steep slopes, when combined with active tectonic processes and seasonal variations in precipitation, create highly favorable conditions for landslides. The 2023 earthquake sequence and subsequent storms triggered more than 3,600 landslides, as reported in preliminary assessments (Görüm et al., 2025, 2023).

**3. Data and Methods**

We followed a five-step methodology comprising: (i) generating surface deformation data, (ii) filtering to highlight localized deformation, (iii) identifying slow-moving landslides, (iv) manually validating and assigning confidence levels based on surface expressions indicative of slow-moving landslides, and (v) analyzing their spatial distribution. The core elements of this approach are illustrated in **Figure 2**.

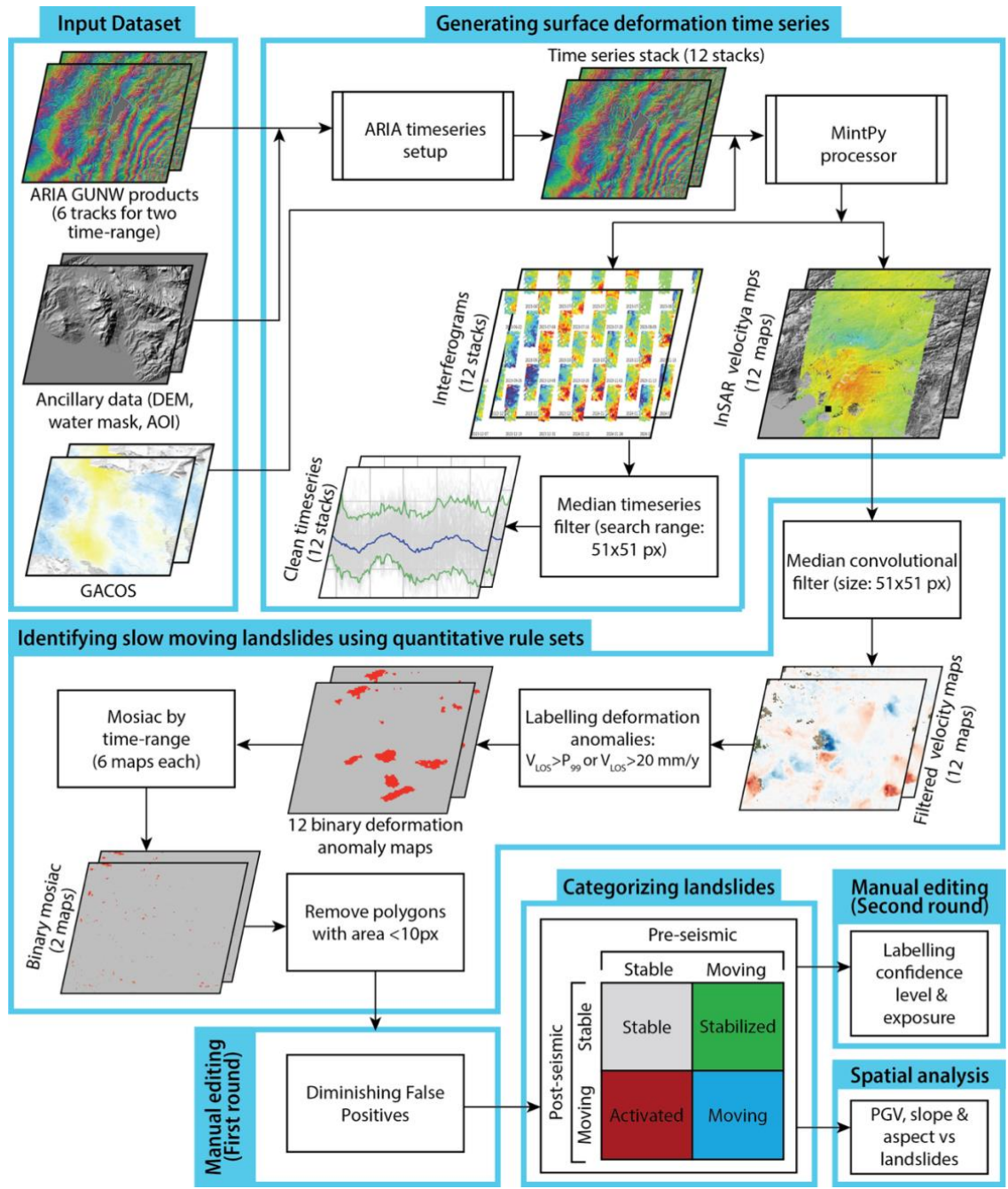

*Figure 2.* Flowchart illustrating the overall five-step methodology used in this study, including time series generation, signal refinement, landslide detection and labeling, validation and spatial distribution analysis.

### 3.1. Generating surface deformation time series

InSAR data were acquired from the ARIA project (Bekaert et al., 2019; Buzzanga et al., 2020). The study region is covered by three ascending tracks (14, 116, and 43) and three descending tracks (94, 21, and 123), as illustrated in **Figure 2**. To construct a meaningful temporal stack for landslide deformation analysis, we retrieved interferograms spanning a period of three years prior to the February 2023 earthquake sequence and one year following it. This temporal selection was deliberately chosen to exclude coseismic deformation effects from the January 2020 Elazığ earthquake (Cakir et al., 2023), thereby isolating the deformation signals relevant to long-term and post-seismic slope behavior. The interferograms are available at a spatial resolution of 3 arc-seconds, corresponding to approximately 90 meters, which supports regional-scale analysis while maintaining reasonable spatial detail.

For the study area defined in the previous section and for each of the six Sentinel-1 tracks, we created time series stacks representing both pre- and post-seismic conditions. To do this, we first downloaded pre-processed Geocoded UNWrapped Interferograms (GUNW) provided by the ARIA project and prepared them for time series processing using the ARIA-Tools (Bekaert et al., 2019; Buzzanga et al., 2020).

For the pre-seismic stage, interferograms were downloaded from February 2020 to early February 2023. For the post-seismic period, data were acquired from late February 2023 to the end of February 2024. The entire month of February 2023 was deliberately excluded to avoid capturing large deformations associated with the mainshock events and major aftershocks.

Once the interferograms were downloaded for each track, we generated time series stacks for both the pre- and post-seismic periods using the ariaTSsetup workflow in ARIA-Tools, resulting in a total of 12 stacks. This workflow automatically identifies the required interferogram products, crops, and stitches them together to produce seamless interferograms. Additionally, water bodies are masked out, and relevant metadata from the downloaded products are extracted to support time series analysis. The extracted metadata includes parameters such as perpendicular and parallel baselines, azimuth angle, incidence angle, and look angle.

From the time series stacks, all interferograms with a temporal baseline exceeding 400 days or a perpendicular baseline greater than 1 km were excluded. Additionally, some interferograms were manually removed based on visual inspection to ensure the continuity and integrity of the interferogram network. Interferogram pairs with an average spatial coherence below a threshold of 0.6 were also discarded. For each stack in the pre-seismic setting, we have at least 177 interferograms with a maximum of 647 interferograms. For the post seismic setting we used at least 78 and 88 interferograms at most. In the temporal baseline, for the pre-seismic stacks, there were interferogram pairs with 6, 12, 18 (12, 24, 36 days after failure of Sentinel 1B satellite) days and a yearlong-interferograms, if their average spatial coherence exceeds the defined threshold. For the post-seismic case, however, yearlong interferograms were not used because of rapid deformation and a small temporal range. In this case, there were interferogram pairs with 12, 24, and 36 days of temporal baseline. The final network for each track with accepted and rejected networks are plotted in the supplementary materials, Figures S1 and S2.

For each track, a reference point was selected based on high temporal coherence and representative terrain characteristics. Once a reliable interferogram network was established, we applied the Small Baseline Subset (SBAS) approach (Berardino et al., 2002) to generate deformation time series using the MintPy software. This process involved correcting unwrapping errors by accounting for bridging and phase closure (Yunjun et al., 2019) and performing network inversion. As the inversion step is particularly time-consuming and computationally intensive, it was executed on the Dutch national supercomputer Snellius, utilizing 150 cores in parallel to enhance computational efficiency.

The inverted time series data often contains noise from various sources, including solid Earth tides, tropospheric delays, orbital ramps, topographic effects, and residual phase errors. While it is not possible to eliminate all of these artifacts, several correction strategies were applied to minimize their impact. **Table 1** summarizes the specific correction methods implemented for each type of noise, along with relevant references from literature.

After applying these corrections, a linear regression was performed on the time series data to estimate the annual deformation velocity for each pixel.

*Table 1. Summary of error correction strategies applied to MintPy time series data.*

| Noise type | Applied Correction | Reference |
|---|---|---|
| Solid earth tides | Corrects the solid Earth tides due to the gravity pulled from the Sun and the Moon. | (Milbert, 2018; Yunjun et al., 2022) |
| Tropospheric delay | Corrects tropospheric delay using the iterative tropospheric decomposition model. | (Yu et al., 2018) |
| Ramp effects | Removes a ramp due to residual long-wavelength interferometric phase components from each acquisition using linear fit. | (Milbert, 2018; Yunjun et al., 2022) |
| Topographic effects | Estimates and removes residual topographic effects (due to DEM errors) which correlate with temporal variation of perpendicular baseline. | (Fattahi and Amelung, 2013) |
| Phase Errors | Estimates and corrects the average noise level for each acquisition by calculating the RMS of the residual phase. | (Yunjun et al., 2019) |

## 3.2. Filtering data to highlight local deformation

The obtained velocity field contains two dominant signals, particularly in the post-seismic period. The first is a long-wavelength deformation component associated with post-seismic viscoelastic relaxation and broader crustal adjustment. The second consists of localized surface deformation resulting from slow-moving landslides and other small-scale near-surface processes. Since the primary objective of this study is to detect and analyze slow-moving landslides, it is essential to decompose the total deformation signal into these two distinct components.

Generally, in geophysical inversion where the long-wavelength deformation (and subsequent velocity) is of interest; high frequency signals are removed using quadtree analysis (Isken, 2017). However, this approach does not suit well for our case. Some previous studies applied the removal of long-wavelength deformation by referencing local manually selected stable points and applying spatial filters (Bekaert et al., 2020). While effective for small areas, this method is labor-intensive and impractical for regional-scale applications.

To overcome this limitation, we developed a convolution-based approach to separate the long-wavelength deformation signal. Specifically, we implemented a sliding window of size n × n, within which the median deformation value is computed. This median is assumed to represent the long-wavelength component within the local window. The optimal window size (n) was determined through iterative testing, with the best results obtained using a 51-pixel window (~5 km); where 25 pixels are present in 4 different sides of the pixel in consideration. Once this median deformation (hereafter referred to as the reference value) is calculated, it is subtracted from the deformation value of the central pixel. This process is repeated across all pixels in the study area to extract the localized deformation component.

We choose the median instead of the mean to minimize the influence of extreme values (e.g., as landslides) that could bias the reference value and eventually reduce the total deformation. Similarly, we did not choose the minimum value because it may represent the deformation which was cancelled due to two distinct deformation processes. This pixel-wise referencing process was applied to all the pixels in velocity map for all the tracks throughout the study area to obtain the cleaner high frequency velocity signal.

**3.3. Identifying slow moving landslides using quantitative rule sets**

Using these corrected velocity maps, we proceeded to identify slow-moving landslides on non-flat areas. To remove flat areas, we masked hillslopes where steepness is less than 5 degrees.

For each track, with annual absolute Line of Sight (LOS) velocity ($V_{LOS}$) exceeding the 99[th] percentile (p99) of all the pixels were classified as slow-moving landslide grids. To avoid omitting landslides due to high p99 values, we added an additional condition where the pixel is classified as landslides regardless of p99 if it exceeds 20 mm/year.

The 20 mm/year threshold was empirically defined to capture several likely slow-moving landslides that had been identified through field observations. It is important to note, however, that our field investigations were not intended to map all slow-moving landslides in the region. Instead, they focused on identifying geomorphological evidence of active hillslope deformation within a representative subset of the study area. By applying this velocity threshold, we aimed to extrapolate our field-based insights to the broader region.

Once binary deformation masks were generated for each track, we mosaicked the results to produce two composite images representing the pre- and post-seismic conditions. These binary composite maps were then converted into polygons, where each polygon contains a pixel or clusters of pixels. From this set, we deleted the polygons that have "0" (i.e., no potential slow-moving landslide) and kept the locations with attribute "1" which represent potential slow-moving landslides. To reduce false positives, we excluded any polygon smaller than 10 pixels in size (i.e., 0,081 km$^2$). This threshold was selected to minimize the influence of isolated noisy pixels that could otherwise lead to false detection.

### 3.4. Expert-based delineating and labeling of slow-moving landslides

Once the potential landslide locations were identified based on the quantitative rule sets described above, we manually evaluated each case in two steps to reduce false positives and enhance the inventory with additional information.

First, we retained polygons only if they appeared consistently in multiple overlapping InSAR tracks. To assess this, we performed a visual inspection to determine whether the velocity signal within each polygon was distinct from its surroundings. If the signal lacked clear boundaries or was mixed with noisy areas, we discarded the polygon to adopt a conservative approach.

Second, we overlaid the remaining polygons onto high-resolution optical imagery obtained from Google Earth basemaps, as well as drone images collected during previously conducted field surveys (Gorum et al., 2023). This allowed us to further diminish false positives and manually refine the landslide polygons based on observable landscape features, as initial outlines derived from InSAR data may not always correspond to morphologically meaningful extents. During this step, we also enriched the inventory by assigning confidence levels to each polygon, reflecting the likelihood of a visible surface expression associated with slow-moving landslides.

It is important to clarify that slow-moving landslides may not always exhibit surface expression in optical imagery. Therefore, the assigned confidence levels should be interpreted as complementary indicators that indirectly support the presence of deformation rather than confirming it outright. In evaluating confidence levels, we considered three criteria related to the presence of deformation indicators: (i) tension cracks and step-like topography; (ii) landforms known to be susceptible to failure, such as old landslide deposits; and (iii) locations of coseismic landslides, which often correspond to zones of increased susceptibility (Gorum et al., 2023). Based on these criteria, we assigned five confidence levels: very low, low, moderate, high, and very high.

As part of the visual inspection of optical imagery, we also carried out an exposure analysis, identifying elements at risk within the landslide polygons. These included settlements, agricultural fields, roads, and reservoirs. Once the landslide polygons were finalized, we classified all hillslopes based on their temporal behavior as follows: hillslopes were considered Stable if no slow-moving landslide polygons were present in either the pre- or post-seismic periods; Stabilized if polygons were present only in the pre-seismic period; Activated if polygons were absent in the pre-seismic but present in the post-seismic period; and Moving if overlapping polygons were present in both periods, regardless of size.

Also, to obtain a general picture of the deformation trend for the landslides activated by the earthquake sequence, we applied a referencing procedure to the time series data similar to the one used for the velocity signal. We then performed a timeseries signal decomposition (van der Walt & Millman, 2010) with a timestep of 15 observations to identify the general deformation trend, which are then plotted with their median, 5th, and 95th percentiles to understand the general deformation behavior.

### 3.5. Examining the spatial distribution of slow-moving landslides

Following the classification, we analyzed the spatial distribution of the landslides with respect to several factors that may influence landslide occurrence, including peak ground velocity (PGV), slope angle, and aspect. We obtained PGV information from the USGS Earthquake Hazards Program (https://earthquake.usgs.gov/storymap/index-Türkiye2023.html). We extracted the maximum recorded PGV values associated with the Pazarcık and Elbistan earthquakes, combining them to represent the seismic forcing across the study area. To generate the topographic variables (slope and aspect), we use the SRTM digital elevation model.

Mean value of the each of these parameters were extracted at the landslide location by zonal statstics operation where the zone is defined by the polygon geometry of manually delineated landslides. For each class of the landslide and factors controlling them we performed Mann-Whitney U test to understand whether their distribution are statistically different or not. Mann-Whitney U test is a non-parametric statistical test used to determine whether there is a significant difference in the distribution of a continuous or ordinal variable between two independent groups. Unlike the t-test, it does not assume that the data are normally distributed, making it suitable for skewed or non-normal data; such as bi-modal distribution. In this context, it is used to compare the distributions of variables such as PGV, slope, and aspect between different classes of slow-moving landslides (namely, Activated, Moving, and Stabilized). A low p-value (typically less than 0.05) indicates that the two groups are likely to come from different distributions.

### 4. Results

We first derived the raw annual velocity maps for both pre- and post-seismic periods (**Figure 3**). Results show that in the pre-seismic setting, the deformation velocities are relatively low and do not exhibit a clear spatial pattern associated with slow-moving landslides (**Figure 3a** and **Figure 3b**). In contrast, the post-seismic maps display significantly higher velocities, with a dominant long-wavelength signal (**Figure 3c** and **Figure 3d**). This post-seismic deformation is primarily attributed to viscoelastic relaxation and afterslip following the main rupture (Barbot and Fialko, 2010).

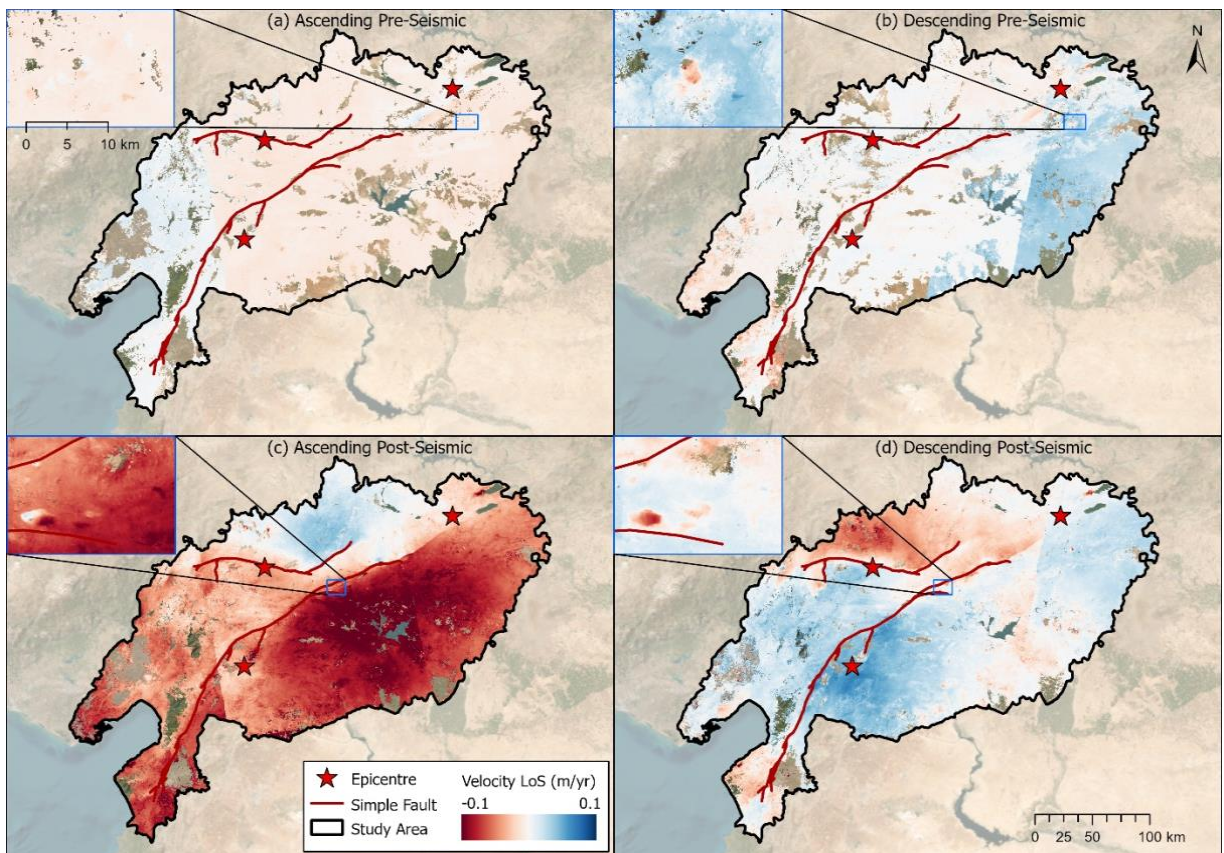

*Figure 3. Raw surface deformation maps generated from ascending and descending Sentinel-1 acquisitions for the (a–b) pre-seismic and (c–d) post-seismic periods. The individual frames are plotted together to represent in a single map but are not merged/stitched.*

Once the long-wavelength signal in the large spatial domain is removed through median filter method, high frequency, small spatial patterns begin to emerge. In **Figure 4**, we can observe such patterns for both pre- (**Figure 4a** and **Figure 4b**) and post-seismic (**Figure 4c** and **Figure 4d**) phases. Even though the overall velocity range is decreased compared to unfiltered conditions (**Figure 3**), where both long- and short-wavelength signals are present, the spatial patterns of the slow-moving landslides are more distinct and interpretable.

To reduce the noise in the dataset and filter out deformation signals which are not prominent, we applied a percentile-based thresholding approach, identifying pixels with $V_{LoS}$ exceeding the 99th percentile as potential indicators of active deformation.

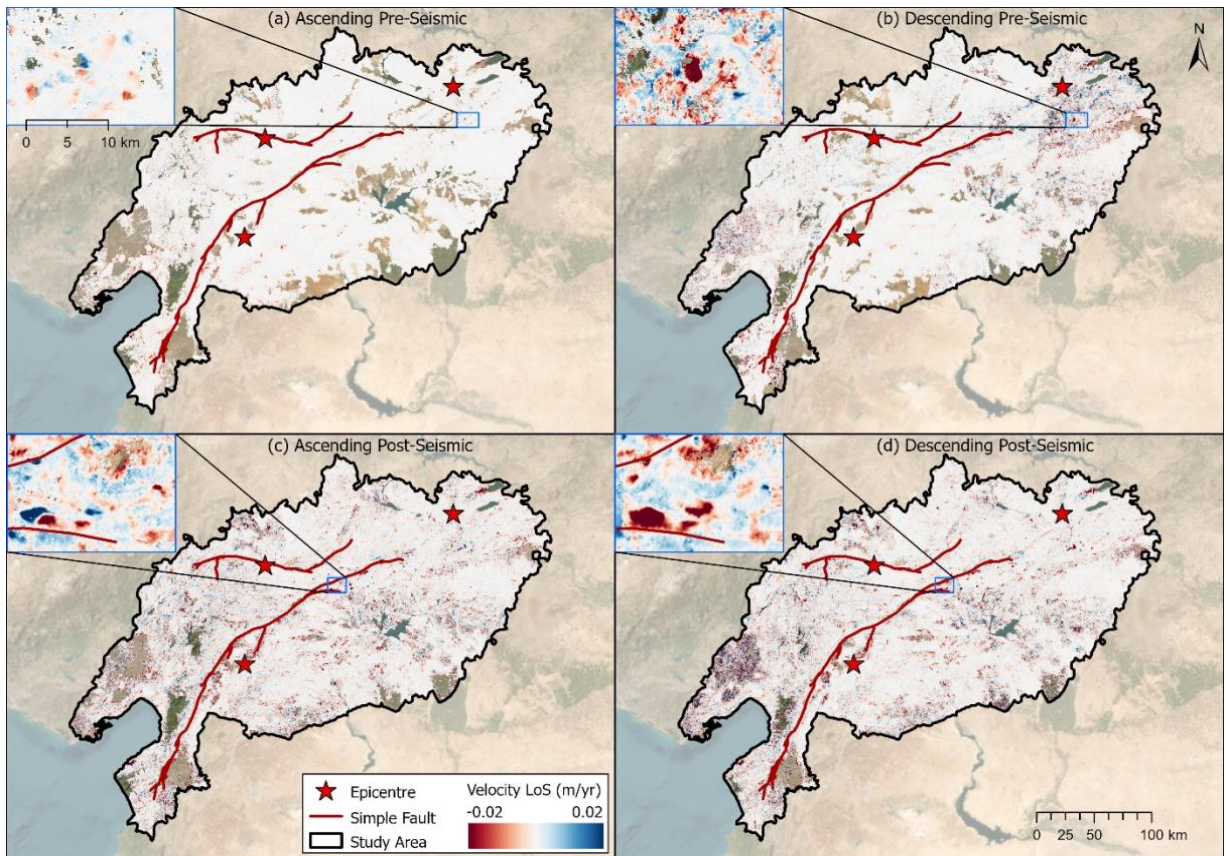

*Figure 4.* Filtered surface deformation maps from ascending and descending orbits for the (a–b) pre-seismic and (c–d) post-seismic periods, after applying median filtering to remove long-wavelength deformation components.

We then manually revised the landslide inventory in two steps. First, we excluded cases where $V_{LOS}$ anomalies were not consistent across different tracks. **Figure 5a** shows an example of an excluded slow-moving landslide polygon, where a $V_{LOS}$ anomaly is visible only in track 43 (**Figure 5b**) but is absent in tracks 116 (**Figure 5c**) and 123 (**Figure 5d**). In contrast, **Figure 5e** presents a retained landslide example, where deformation anomalies are consistently observed across all tracks (**Figure 5f–h**).

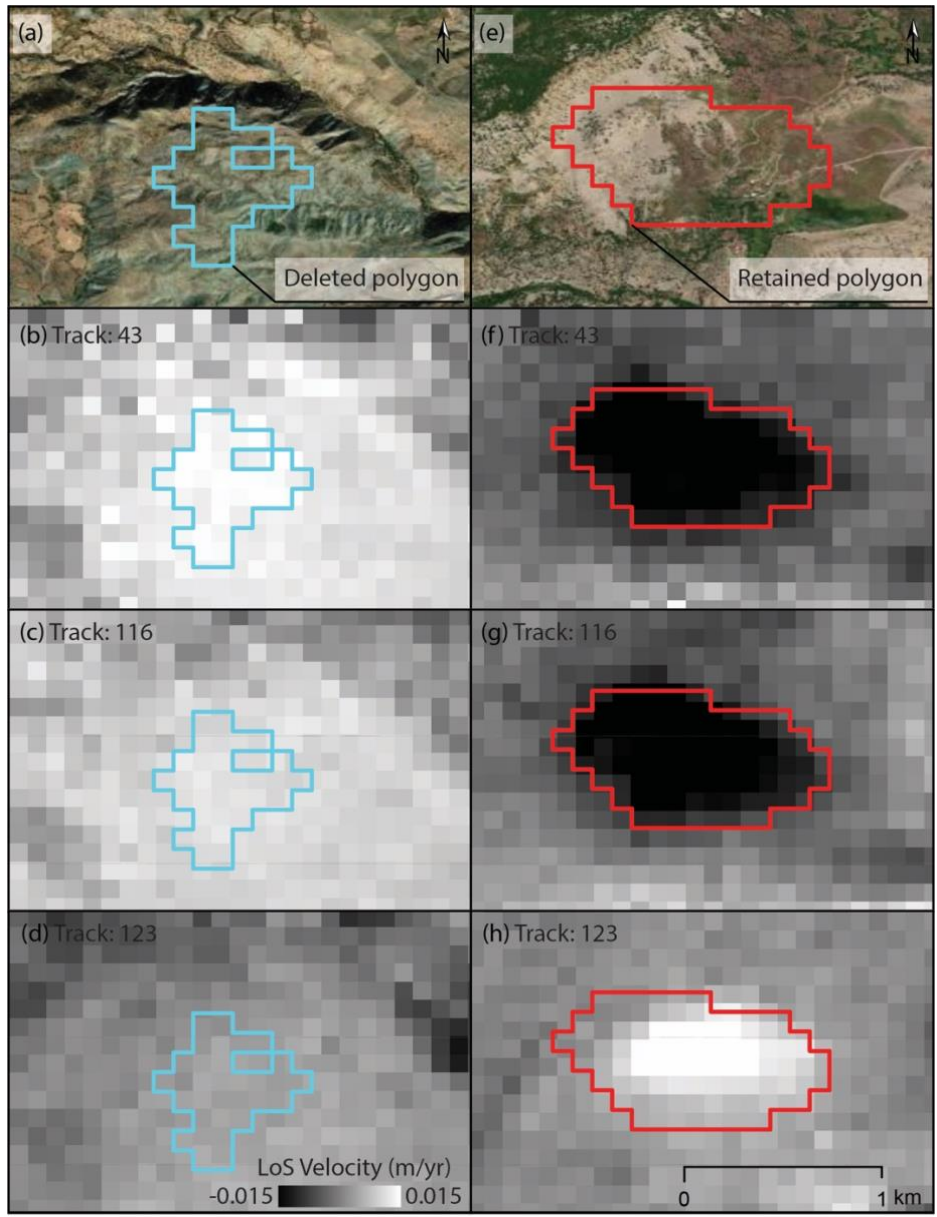

*Figure 5. Example illustrating the first step of the manual editing procedure for the slow-moving landslide inventory, based on consistency across $V_{LOS}$ maps generated from SAR data acquired along different tracks. Panels (a) and (b) show examples of deleted and retained polygons, respectively, overlaid on Google Earth imagery. $V_{LOS}$ maps from multiple tracks are shown in Panels (c–d) for the deleted polygon and in Panels (f–h) for the retained polygon.*

Second, we reviewed optical imagery and landforms using Google Earth to further refine the inventory. **Figure 6** shows an example of this process, where vectorized polygons generated based on quantitative rule sets were manually converted into slow-moving landslide polygons. During this step, we not only delineated morphologically meaningful landslide boundaries (**Figure 6b**) but also removed polygons that appeared to be false positives. For instance, the rectangular area enclosing the highest portion of a gently sloping landscape in **Figure 6a** was filtered out during manual mapping (**Figure 6b**).

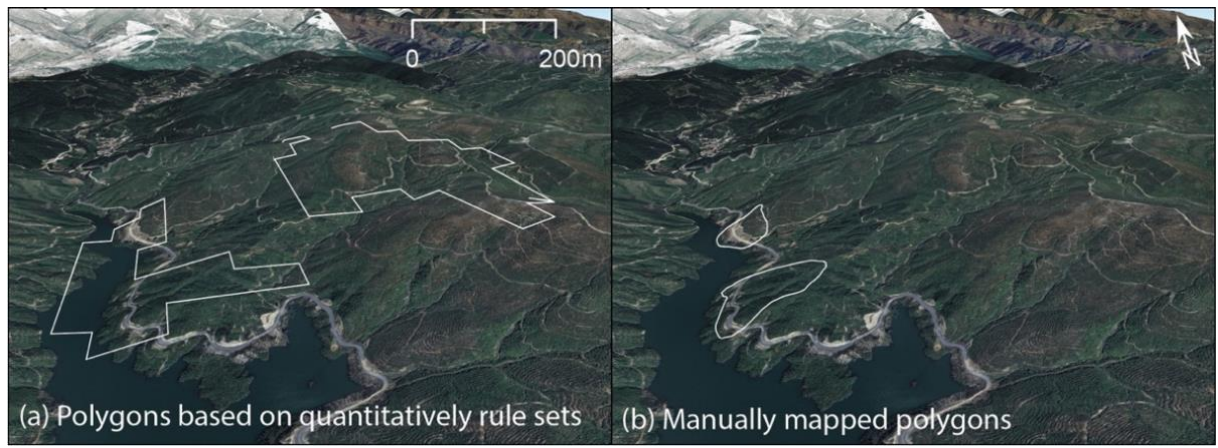

*Figure 6. Example illustrating the second step of the manual editing procedure. Where, (a) vectorized landslide polygons (lat: 36.20° and lon: 35.98°) were converted into (b) morphologically meaningful units through visual inspection of optical imagery and landform features using Google Earth.*

After polygonising and minimizing the false positives from the landslide inventory, we have observed that in the pre-seismic phase, there were total of 48 slow-moving landslides and in the post-seismic period, it increased to 368; resulting in a 7.6-fold increase in total number of landslides (**Figure 7**). While looking at their spatial distribution, 348 landslides were classified as Activated and 20 were classified as Moving whereas 28 were classified as Stabilized/Failed.

In terms of spatial extent, the total slow-moving landslide affected area expanded from 12.36 km² in the pre-seismic phase to 159.39 km² in the post-seismic phase, with a 12.89-fold increase. In each class, it was observed that hillslopes with an area of 131.14 km² were activated following the earthquake, exhibiting new hillslope deformation. An area of 4.61 km², which had shown movement during the pre-seismic phase, became stable afterward. Additionally, the area classified as moving in the pre-seismic period expanded from 7.75 km² to 28.25 km² in the post-seismic period. This indicates that the overall slow-moving landslide budget increased by a factor of 12.9 from the pre- to post-seismic phases, with the total affected area expanding from 12.35 km² to 159.384 km² (**Figure 7a**).

To examine the relationship between the classified landslides and key controlling geomorphic and seismic factors, we analyzed their distribution in relation to PGV, slope steepness (**Figure 7b**), and hillslope aspect (**Figure 7c**). The results indicate that slow-moving landslides classified as Moving exhibit slightly different behavior compared to the Activated and Stabilized categories. The Moving landslides are primarily located in areas subjected to less intense ground shaking (with a peak PGV of 18 cm/s) and on gentler slopes (with a median steepness of 14 degrees). In contrast, the Activated and Stabilized landslides are associated with higher median PGV values of around 32 cm/s and steeper slopes, with a median of approximately 17 degrees.

The distribution of hillslope aspect shows a similar pattern across all categories; slow-moving landslides are mostly concentrated on east- and west-facing slopes, while relatively fewer are located on south-facing slopes. Therefore, we chose to present the frequency distribution collectively for all types, rather than displaying each category separately (**Figure 7a**).

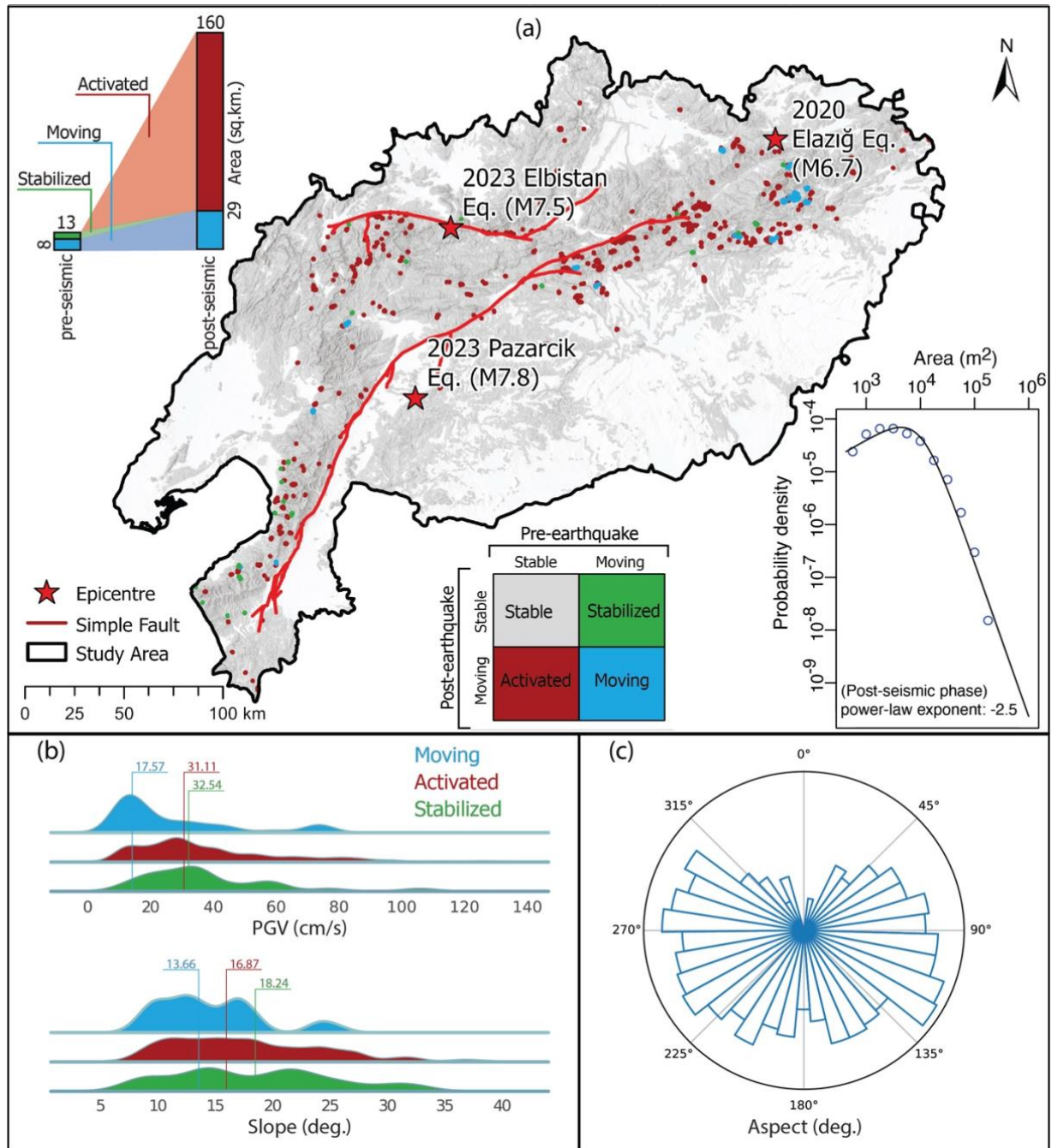

*Figure 7. Plots showing the distribution of slow-moving landslides in terms of: (a) spatial distribution of identified landslides; (b) frequency distributions of peak ground velocity (PGV) and slope steepness for different landslide categories (i.e., Moving, Activated, and Stabilized); (c) aspect distribution of all slow-moving landslides.*

We also observed that the Stabilized landslides geologically diverge from two other categories. Sedimentary rocks appear to be the dominant geology associated with all categories, while ophiolitic units are the second most dominant one only in the Stabilized category (**Figure S3**).

While performing Mann-Whitney U to check whether each of the landslide classes has different distributions of variables. We identified that for the Moving class had significantly ($p<0.05$) different distribution to Stabilized and Activated class. Similarly, between Stabilized and Activated classes, there was no significantly different distribution of the variables.

We analyzed the size distribution of post-seismic slow-moving landslides (**Figure 7a**) and fitted a curve using the double-Pareto simplified distribution (Rossi et al., 2012). The landslide sizes range from 0.01 km² to 7.31 km², with a calculated power-law exponent of -2.5, which matches the global average reported in coseismic landslide inventories (Tanyaş et al., 2018). This suggests that the size distribution of slow-moving landslides does not significantly differ from that of rapidly occurring coseismic landslides.

We also assigned a confidence level to each slow-moving landslide polygon and described the corresponding elements exposed to landslides, as outlined in the Methods section. **Figure 8** presents examples illustrating this expert-based step, aimed at enriching the dataset.

For instance, when no supporting evidence was found based on the evaluation criteria, we assigned a very low confidence level to the polygon. **Figure 8a** illustrates such a case, where the delineated slow-moving landslide polygon is located on a gentle, bare hillslope composed of massive limestone units, with no visible signs of surface deformation or evidence of previous landslides.

Conversely, when clear indications of surface deformation were observed (i.e., the first criterion), we assigned a very high confidence level. **Figure 8b** presents such a case, where the crown of the landslide has already been illuminated due to movement, and tension cracks are visible in the upper part of the hillslope. This movement occurs in a remote area, where no man-made structures are exposed to the landslide, but forested land is affected. This is also reported in the inventory.

In some cases, when deformation anomalies were observed on paleo-landslide bodies (i.e., the second criterion), we assigned a relatively high confidence level, as the old landslide deposits may have been reactivated. **Figure 8c** and **Figure 8d** correspond to such examples. The former is located on a relatively steep slope and poses a threat to a nearby settlement. The latter, on the other hand, is on a gentler hillslope but threatens not only a settlement but also a main road and a reservoir.

Finally, **Figure 8e** provides an example where a high confidence level was assigned based on similarities with co-seismic landslide locations (i.e., the third criterion). In the southern part of the earthquake-affected region, many translational slides occurred during the event on gentle hillslopes (**Figure 8f**), which are underlain by gently tilted sedimentary units (Gorum et al., 2023).

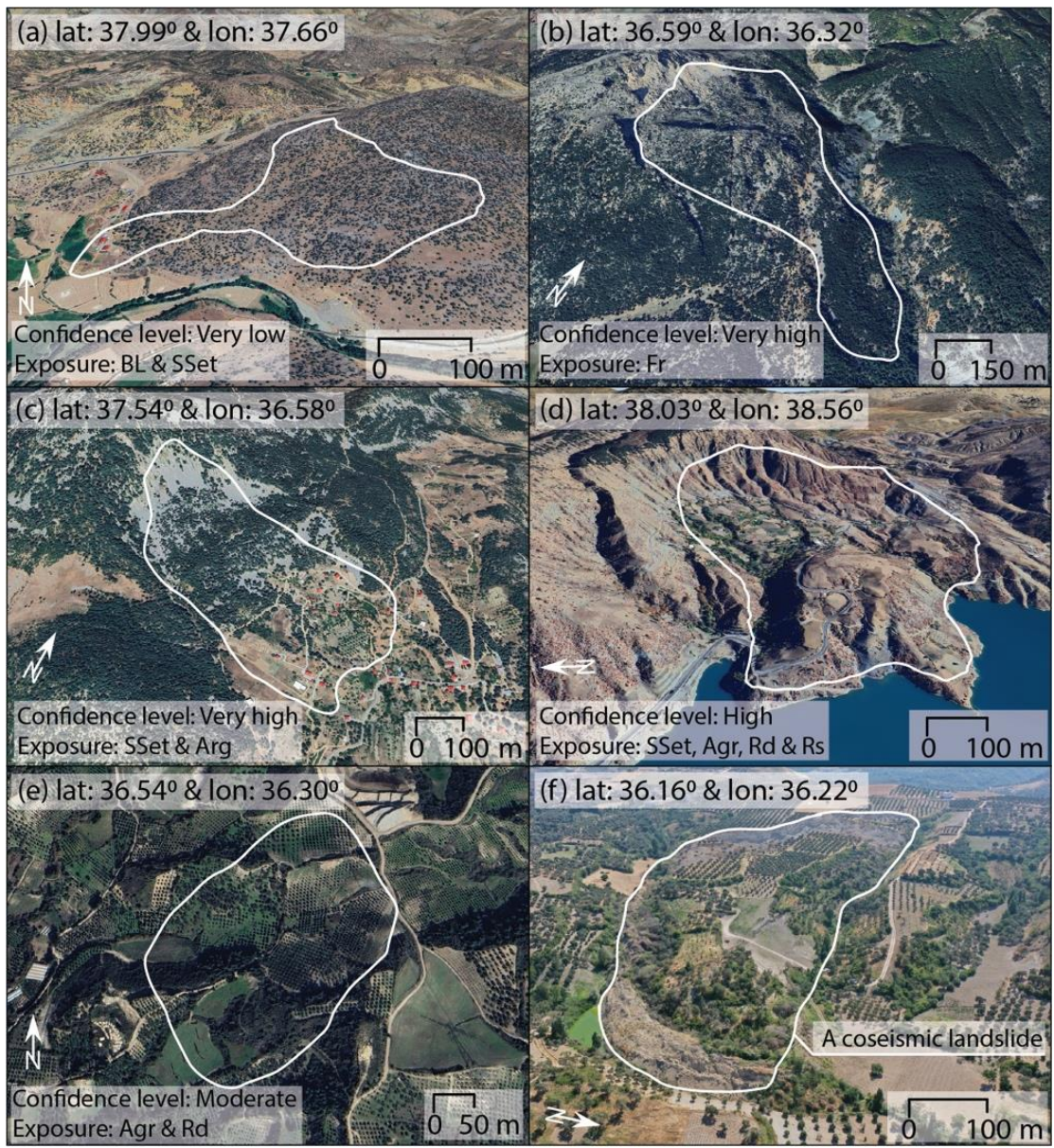

*Figure 8.* Examples illustrating the labeling of slow-moving landslides with respect to confidence level and exposure. (a) very low confidence, with exposure of bare land (BL) and a small settlement (SSet); (b) very high confidence, with exposure of forest (Fr); (c) very high confidence, with exposure of SSet and agricultural area (Agr); (d) high confidence, with exposure of SSet, Agr, road (Rd), and dam reservoir (Rs); and (e) moderate confidence, with exposure of Agr and Rd. Panel (f) shows a coseismic landslide location, suggesting that similar deformation may also occur in areas like the one shown in panel (e).

We also examined and validated some slow-moving landslide locations through field investigations, including drone surveys, to assess the post-seismic conditions of hillslopes affected by the earthquake sequence. For conciseness, we present two representative examples from the central part of the earthquake-affected region in Adıyaman (**Figure 9**): (a) Cankara, Gölbaşı, and (b) Meryemuşağı, Tut.

The first site (**Figure 9a** and **Figure 9b**) corresponds to a previously occurred landslide that exhibited retrogressive movement during the coseismic phase, due to intense ground shaking where PGA reached 0.6g. Our InSAR analysis shows that this hillslope was stable prior to the earthquake sequence but became active afterward, reaching a mean $V_{LOS}$ of -41 mm/year (**Figure 9c**).

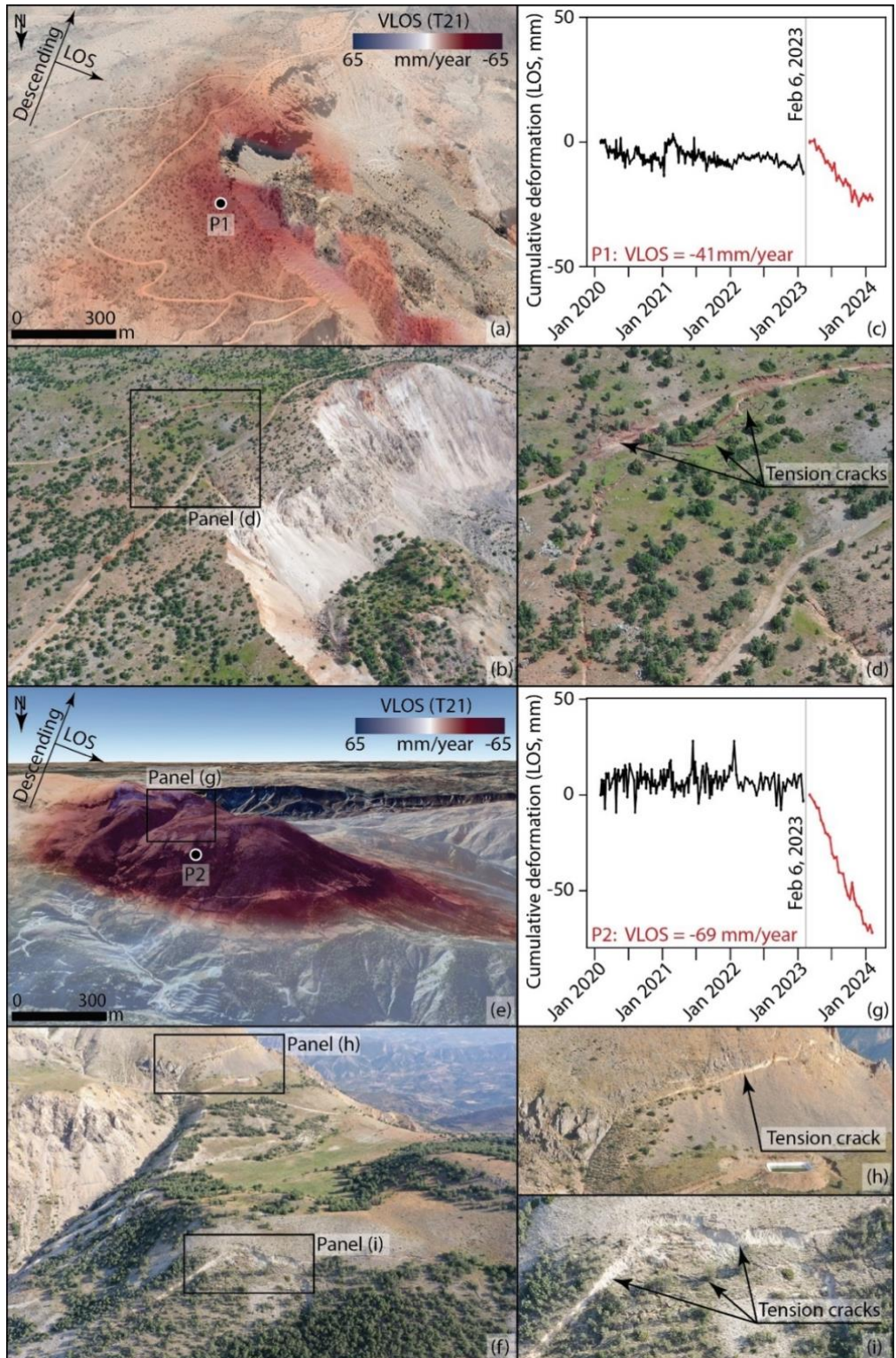

*Figure 9.* Google Earth and drone imagery illustrating field validation of identified slow-moving landslide location. (a–d) Çankara, Gölbaşı (lat-lon of P1: 37.87°-37.86°), and (e–i) Meryemuşağı, Tut (lat-lon of P2: 37.83°-37.87°). Panel (a) shows a satellite image of the hillslope overlaid with deformation rates, where $V_{LOS}$ reaches up to ~50 mm/year near the crown of the landslide at a selected point (P1). Panels (c–d) present the deformation time series for this point, while panel (d) also highlights tension cracks generated during the coseismic event. Panels (e–i) show similar observations for a hillslope in Meryemuşağı, Tut.

The second site (**Figure 9e**) is a large hillslope (~4 km²) that had shown no signs of instability before the earthquake sequence (**Figure 9g**) and had no known history of significant landsliding. However, tension cracks (**Figure 9f**, **9h**, and **9i**) were generated due to the shaking (PGA = 0.4g), and post-seismic surface deformation reached a mean $V_{LOS}$ of -69 mm/year (**Figure 9g**). In both cases, the presence of tension cracks observed in Google Earth imagery and drone photographs supports our InSAR-based identification of slow-moving landslides.

While **Figure 9** presents specific examples of deformation time series and slow-moving landslides, we also aimed to highlight broader deformation trends across the study area. We analyzed both ascending and descending tracks, plotted aggregated time series, and created box plots of absolute velocities for the Activated, Moving, and Stabilized landslide categories (**Figure 10**).

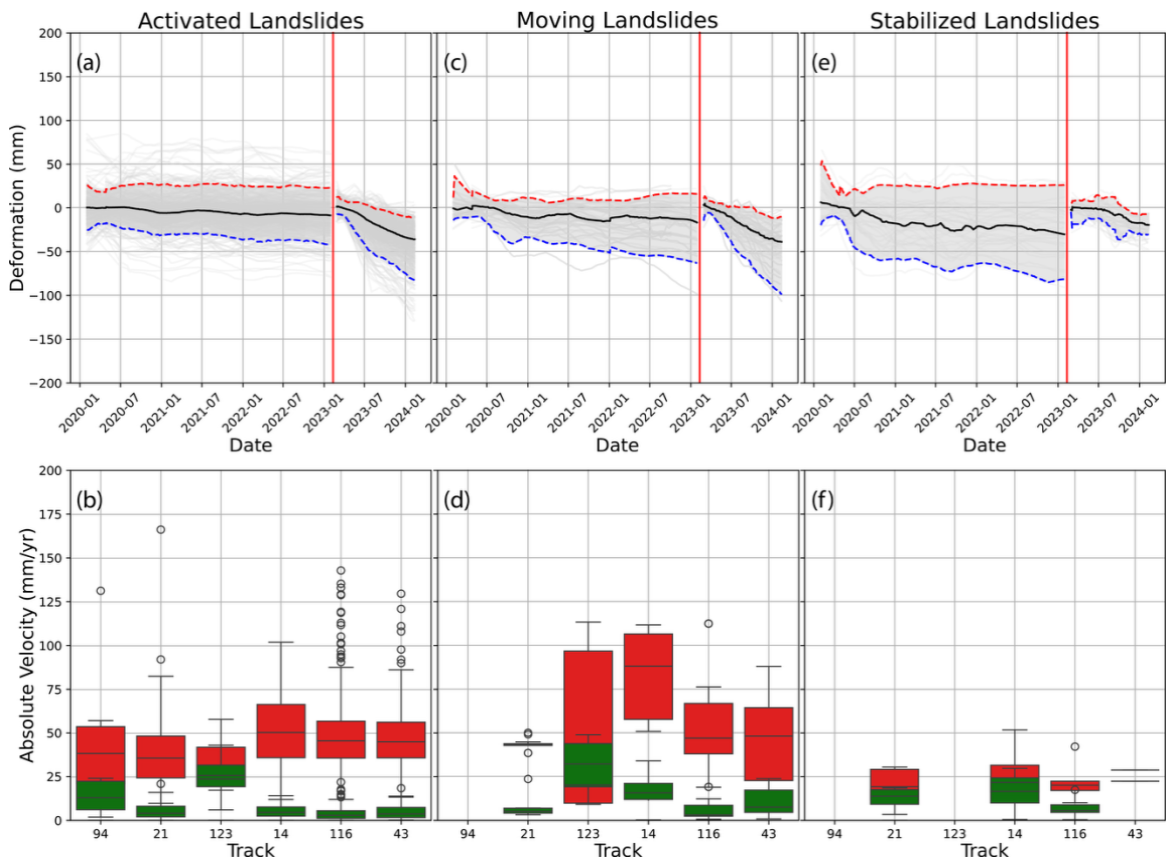

*Figure 10.* Line-of-sight surface deformation time series and absolute velocities, shown as box plots. (a, b) Activated landslides, (c, d) Moving landslides, and (e, f) Stabilized landslides. For simplicity, deformation time series are shown only for locations with negative velocities. In panels (a), (c), and (e), individual landslide deformation time series are represented by grey lines, while the mean behavior per track is indicated by colored lines. Red and blue dashed lines represent the mean plus and minus one standard deviation, respectively, and the solid black line indicates the overall mean trend.

**Figure 10a** shows a sudden and pronounced shift in the deformation trend of the Activated landslides following the earthquake, indicating an increase in absolute velocity across all tracks (**Figure 10b**). The colored mean curves for each track clearly show that these locations were largely inactive prior to the earthquake and began deforming afterwards.

For the Moving category, a similar pattern of active post-seismic deformation is evident in the time series (**Figure 10c**). However, in this case, slight pre-seismic deformation is also observed, which is reflected in the absolute mean velocities shown in the box plots (**Figure 10d**). These results suggest that landslides in the Moving category mostly accelerated following the earthquake sequence.

Finally, the Stabilized landslides show a decreasing deformation trend after the earthquakes (**Figure 10e**), which is also reflected in their lower post-seismic velocities compared to the other categories (**Figure 10f**). Notably, **Figure 10f** also indicates that fewer time series were available for the stabilized landslides, as coherent observations were lacking in many tracks.

## 5. Discussion

With respect to the dual effect of large earthquakes on the slow-moving landslide budget of a landscape, our findings indicate a dramatic increase following the 2023 earthquake sequence. This pattern does not align with observations from the San Andreas Fault system, (Scheingross et al., 2013) or the region affected by the 2008 Wenchuan earthquake (Lu et al., 2025), where relatively few slow-moving landslides were reported near earthquake source areas. However, this difference should not be considered contradictory, as each of these landscapes is shaped by distinct morphometric, climatic, and seismic conditions. Therefore, the observations should be interpreted as site-specific.

It is important to note that compiling reliable slow-moving landslide inventories at regional scales is always challenging. The studies focusing on the San Andreas Fault system, the Wenchuan earthquake region, and our own research all face similar limitations, where the associated uncertainties must be taken into account. To reduce such uncertainties, we combined a quantitative, rule-based landslide identification approach with a two-step expert-based evaluation process. As a result, we assigned a confidence level to each slow-moving landslide polygon, ranging from very low to very high.

When the entire dataset is considered without filtering by confidence level, we observe a 7.6-fold increase in the total number of slow-moving landslides, rising from 48 in the pre-seismic phase to 368 in the post-seismic phase (**Figure S4**). In contrast, when only high and very high confidence observations are retained, the number of slow-moving landslides increases from 5 in the pre-seismic period to 50 in the post-seismic period, indicating a 10-fold increase. In this case, the counts for Activated, Moving, and Stabilized landslides drop from 348 to 46, 20 to 4, and 28 to 1, respectively.

The spatial distribution of these high-confidence slow-moving landslides also reveals a distinct pattern: most of the Moving category landslides are concentrated in the northeastern part of the study area, while the Activated landslides tend to align along the fault trace (**Figure S4**).

Among the three categories, the Stabilized landslides may be associated either with rapid failures that occurred during the coseismic event or with a combination of geomorphological and geomechanical processes that typically follow major earthquakes.

Former mechanism refers to the reconfiguration of slope geometry following initial failure; during strong ground shaking, unstable slopes often experience partial or full collapse, resulting in a new topographic equilibrium that is more stable under prevailing conditions (Lv et al., 2025; Song et al., 2022). In many cases, this process removes critical overburden stress or steep slope angles, leading to a mechanically more stable configuration. However, based on the preliminary coseismic landslide (Görüm et al., 2023) and our own inspections of optical imagery, we did not observe any landslide activity at these locations. This suggests that the latter explanation is more plausible and that it is associated with the physical consolidation of slope materials. The intense seismic loading can densify loose, cohesionless soils and debris, reducing their susceptibility to further deformation. This densification can result from cyclic shaking that causes the rearrangement of soil particles into a more compact structure, thereby increasing shear strength and reducing pore space (Lv et al., 2025; Song et al., 2022).

We should stress that ophiolitic units, which are composed of highly weathered and landslide-prone materials such as schist, gneiss, and pelagic limestone, are present in approximately 40 percent of the stabilized landslide locations. Notably, the only stabilized slow-moving landslide with high confidence is also located within these ophiolitic units, which may suggest that densification processes, as previously described, contributed to its stabilization.

A similar process may also partly explain the decelerating trends observed in the deformation time series of the Activated landslides. Although an overall increase in deformation is evident following the 2023 earthquake sequence, this trend begins to moderate by early 2024, with a noticeable decline in deformation rates at many observed locations. This deceleration suggests that slow-moving landslides in the region do not exhibit a simple linear response to seismic forcing but rather follow a complex temporal evolution, often transitioning into a quasi-stable state after a phase of accelerated movement.

One of the primary drivers behind this deceleration could be the redistribution and relaxation of stress following the mainshock. Earthquakes generate abrupt changes in the local stress regime, temporarily increasing the shear stress acting on slope materials and initiating motion. Over time, as the crust re-equilibrates, these stress anomalies dissipate, reducing the driving forces behind continued slope instability. Similar behavior was documented by Lacroix et al. (2014), who observed a landslide that experienced coseismic displacement followed by post-seismic slip three times larger than the initial movement. This post-seismic slip gradually decayed over approximately five weeks, eventually returning to the pre-event velocity, a pattern they successfully modeled using a rate-and-state friction framework. Their findings suggest that landslides, like tectonic faults, can exhibit afterslip behavior followed by stabilization, which may explain the temporal evolution seen in our Activated landslide category.

Another factor involved might be the mechanical recovery of slope materials under certain conditions. During rapid movement phases, the internal structure of soils can evolve through strain hardening, densification, or frictional healing, which increases resistance to continued deformation. Bontemps et al. (2020) observed that frictional healing within the basal shear zone leads to a steady reduction in landslide velocities after acceleration, reflecting a progressive increase in shear resistance with time. This mechanical strengthening can enhance the overall shear strength of the material and counteract the forces that initiated the post-seismic acceleration. As a result, landslides may gradually transition into a deceleration phase once a new mechanical configuration is established.

Despite the robust regional-scale characterization achieved through this study, several limitations remain. One of the constraint lies in the spatial resolution of the interferograms generated through the ARIA; smaller size slow-moving landslides may not have been adequately captured as the ~90m pixel size and least pixels required to be classified as landslide may have generated some false negatives. Which could be avoided with the higher-resolution dataset, but which also comes with computational challenges, especially for a large spatial extent. Similarly, the filtering approach that we have used might also create some locations with false positives or false negatives. Mainly if the median value is much larger than the given pixel, it might be false positives. Similarly, if the median value is almost similar to the landslide pixel, then it might create a false negative signal. This can be sensitive to the filter type (e.g. mean, median, minimum); window size and the spatial distribution of deformation itself.

The large-scale processing required to cover the full extent of the study area may introduce signal mixing from non-landslide deformation sources, increasing the potential for misclassification and contributing to uncertainty in the final inventory. These challenges are further compounded by the limited extent of field validation, as conducting ground-based surveys across the entire 66,800 km$^2$ study area is not feasible. Moreover, while this study documents the activation and progression of surface deformation, it does not explore the underlying geotechnical mechanisms. Specifically, the changes in material strength or slope stability parameters that may occur following seismic loading. Understanding these processes remains an interesting direction for future research and requires physical modelling/inversion of hillslope strength characteristics with respect to given deformation signals. Finally, the classification of landslides in this study was performed using a threshold-based approach. While effective for large-scale mapping, future work could benefit from the integration of unsupervised machine learning or data-driven classification techniques, which may offer more nuanced delineation of slow-moving landslide boundaries and better capture the complex variability in deformation behavior.

## 6. Conclusion

This study presents a regional-scale assessment of hillslope deformation in the tectonically active region surrounding the East Anatolian Fault Zone. By analyzing three years of pre-seismic and one year of post-seismic InSAR-derived deformation time series following the 2023 Kahramanmaraş earthquake sequence, we mapped deformation anomalies across approximately 66,800 $km^2$.

We categorized slow-moving landslides based on their temporal evolution between the pre- and post-seismic phases. Our results show that 131.14 $km^2$ of hillslopes began actively deforming after the earthquakes, while 4.62 $km^2$ of terrain stabilized. Meanwhile, areas already deforming before the earthquakes expanded from 7.75 $km^2$ to 28.25 $km^2$.

In total, the area affected by slow-moving landslides increased from 12.36 $km^2$ in the pre-seismic phase to 159.39 $km^2$ in the post-seismic phase, reflecting an increase by a factor of 12.9. These findings indicate that the earthquake sequence significantly increased the number and extent of slow-moving landslides in the region. Although strong earthquakes can have both triggering and stabilizing effects, our results suggest that triggering and reactivation are dominant in this case.

We also found that deformation trends among moving landslides generally accelerated after the 2023 earthquake, followed by a gradual decline in early 2024. This non-linear behavior highlights the importance of long-term monitoring to capture the full progression of slope instability in seismically affected landscapes.

To enhance the value of our InSAR-derived slow-moving landslide inventory, we added expert-based confidence levels for each landslide and performed exposure analyses to identify elements at risk. By sharing our outputs, including the landslide inventory, we aim to support more comprehensive post-seismic landslide hazard assessments. The identified slow-moving landslides may still pose future threats and thus require further investigation and continued monitoring.

In conclusion, this study contributes to a better understanding of slow-moving landslides in tectonically active regions and provides a basis for future research integrating geotechnical assessments, high-resolution datasets, and improved classification techniques.

**Acknowledgements**

We thank Tolga Görüm for his insights regarding the in-situ conditions. Part of this research was carried out at the Jet Propulsion Laboratory, California Institute of Technology, under a contract with the National Aeronautics and Space Administration (80NM0018D0004) and supported by NASA Earth Surface and Interior Program. We acknowledge ESA for processing the Copernicus Sentinel-1 level-1 images available through the Alaska Satellite Facility (ASF) for free, and ARIA for processing the InSAR data and ASF for providing open and free access to the standard InSAR displacement products. This work contains modified Copernicus data from the Sentinel-1A and -1B satellites processed by the ESA. This work was partially funded by the NATO Science for Peace and Security Program (SPS project G6190) and through the ESA contract 4000144311/24/I-KE, AMHEI (Advancing knowledge of Multi-Hazards processes and their Impact). The computation time and resources were provided by the Dutch national e-infrastructure with the support of the SURF Cooperative using grant no. EINF-12354.

https://doi.org/10.1016/J.CAGEO.2019.104331

Yunjun, Z., Fattahi, H., Pi, X., Rosen, P., Simons, M., Agram, P., Aoki, Y., 2022. Range Geolocation Accuracy of C-/L-Band SAR and its Implications for Operational Stack Coregistration. IEEE Trans. Geosci. Remote Sens. 60. https://doi.org/10.1109/TGRS.2022.3168509

**Supplementary Materials**

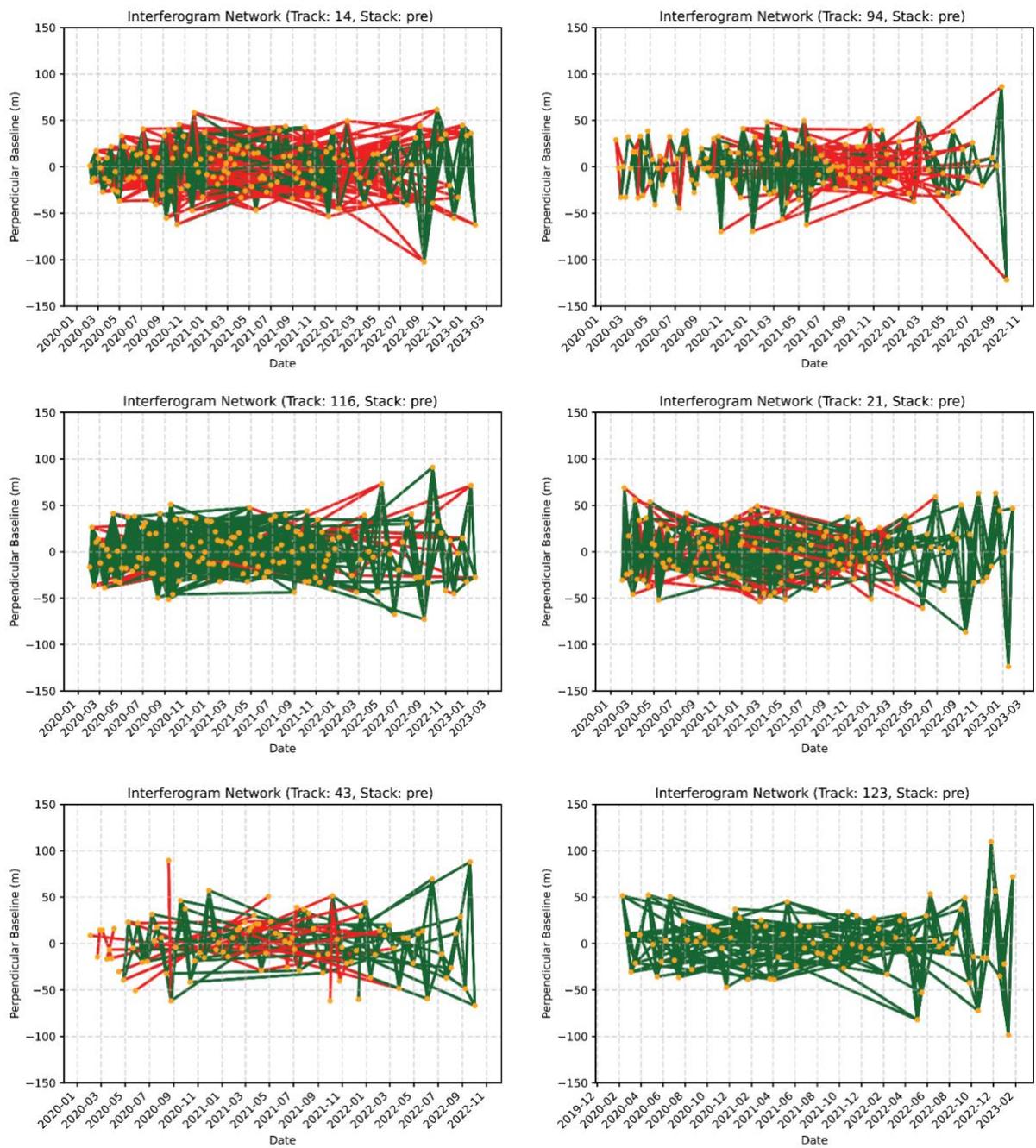

*Figure S 1: InSAR network plot for the pre-seismic stack. The red lines represent dropped interferograms and the green line represents the retained interferograms.*

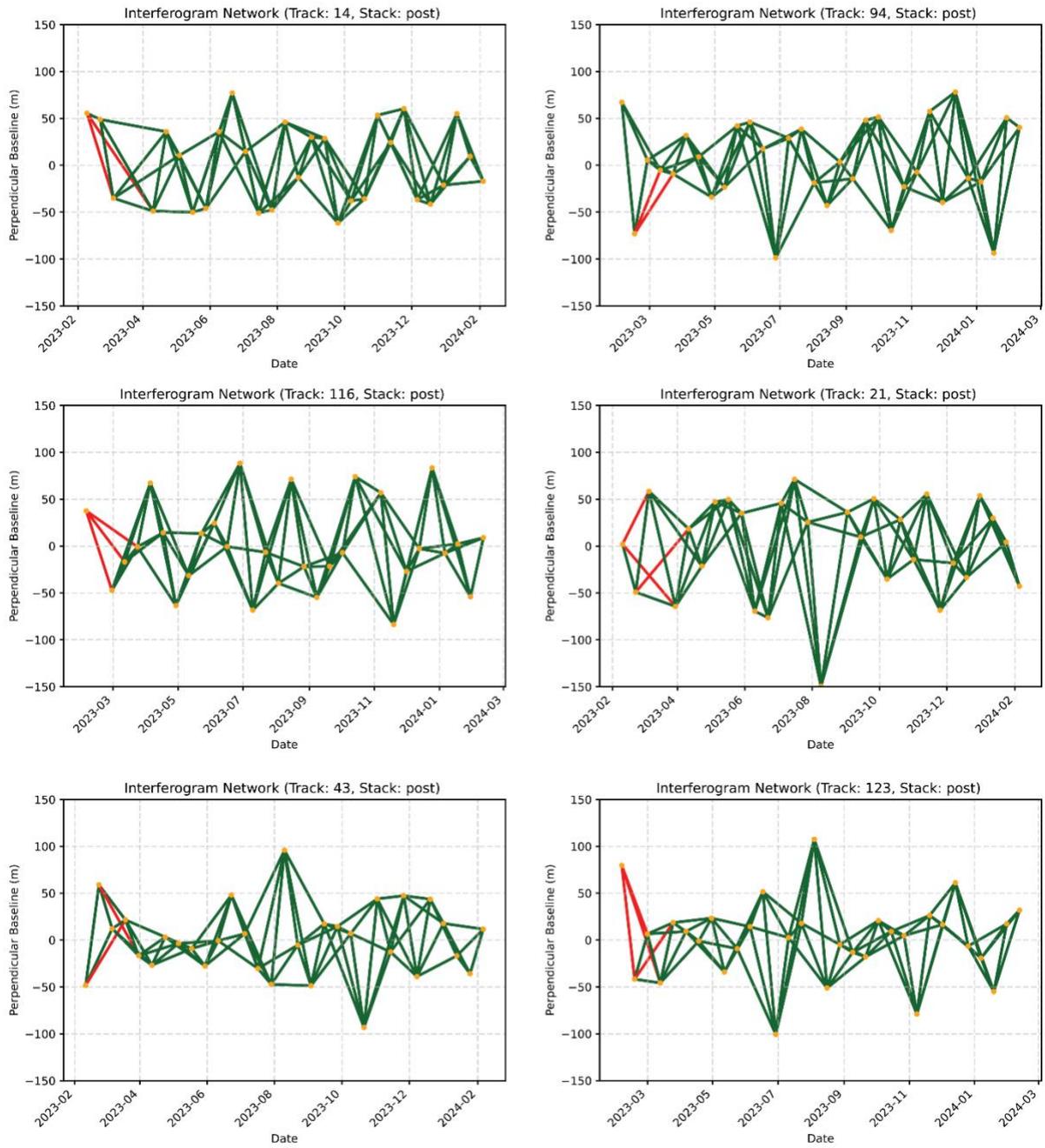

*Figure S 2: InSAR network plot for the post-seismic stack. The red lines represent dropped interferograms and green lines represent the retained interferograms.*

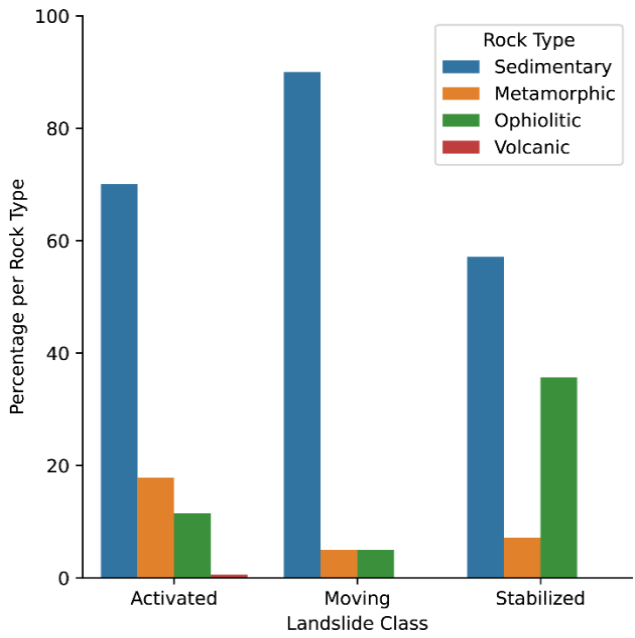

*Figure S 3: Distribution of rock type per calss of the landslide distribution.*

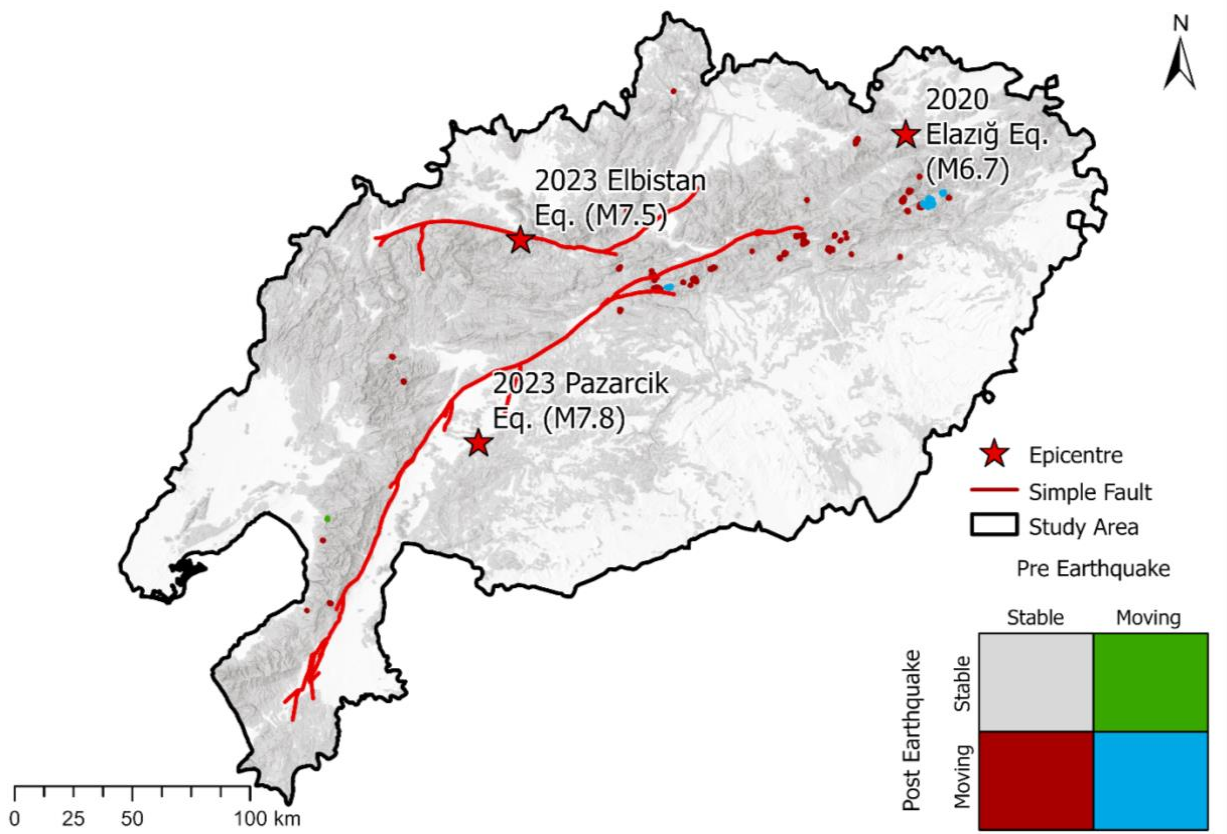

*Figure S 4: Spatial distribution of slow-moving landsldies with high and very high level of confidence*